\documentclass[fleqn,11pt]{wlpeerj}

\usepackage{epstopdf}
\usepackage{graphicx}
\usepackage{multirow}
\usepackage{booktabs}
\usepackage{rotating}

\title{Network connectivity optimization: An evaluation of heuristics applied to complex networks and a transportation case study}

\author[1]{Jeremy Auerbach}
\author[2]{Hyun Kim}
\affil[1]{Colorado State University\\
Department of Environmental and Radiological Health Sciences\\
Fort Collins, CO 80523, USA}
\affil[2]{University of Tennessee\\
Department of Geography\\
Knoxville, TN 37996, USA}
\corrauthor[1]{Jeremy Auerbach}{jeremy.auerbach@colostate.edu}

\begin{abstract}
Network optimization has generally been focused on solving network flow problems, but recently there have been investigations into optimizing network characteristics. Optimizing network connectivity to maximize the number of nodes within a given distance to a focal node and then minimizing the number and length of additional connections has not been as thoroughly explored, yet is important in several domains including transportation planning, telecommunications networks, and geospatial analysis. We compare several heuristics to explore this network connectivity optimization problem with the use of random networks, including the introduction of two planar random networks that are useful for spatial network simulation research, and a real-world case study from urban planning and public health. We observe significant variation between nodal characteristics and optimal connections across network types. This result along with the computational costs of the search for optimal solutions highlights the difficulty of finding effective heuristics. A novel genetic algorithm is proposed and we find this optimization heuristic outperforms existing techniques and describe how it can be applied to other combinatorial and dynamic problems.
\end{abstract}

\begin{document}

\flushbottom
\maketitle
\thispagestyle{empty}

\begin{table}[ht!]
\centering
\tiny
\begin{tabular}{l l}
\toprule
Symbol				& Definition
\\
\midrule
$\nu$				& Network node
\\
$e$					& Network edge
\\
$N$					& Number of nodes in a given network, $N=\sum_i \nu_i$
\\
$A$					& Network adjacency matrix
\\
$a_{ij}$			& Adjacency matrix element $ij$
\\
$F$					& Focal node
\\
$d(i,j)$			& Network distance between nodes $i$ and $j$
\\
$D$					& Threshold distance from focal node
\\
$N_C$				& Set of close nodes, $N_C \subset N$
\\
$N_D$				& Set of distant nodes, $N_D \subset N$
\\
$N_C'$				& Set of nodes that are now close after a new connection
\\
$L_F$				& Average path length to the focal node
\\
\midrule
$C(i,j)$			& Cost of the new connection
\\
$B(i,j)$			& Benefit of the new connection
\\
$\alpha$			& Cost weight
\\
$\beta$				& Benefit weight
\\
$t$					& Optimization iteration
\\
$O^t$				& Optimal solution for iteration $t$
\\
$O^*$				& Optimal solution
\\
$M$					& Set of long-term memory solutions
\\
$N_i^C$				& Set of neighboring close nodes for node $i$
\\
$N_j^D$				& Set of neighboring distant nodes for node $j$
\\
\midrule
$C_i^D$			& Degree centrality of node $i$
\\
$C_i^C$			& Closeness centrality of node $i$
\\
$\sigma_{ij}$			& Shortest path between nodes $i$ and $j$
\\
$\sigma_{jk}(i)$		& Shortest path between nodes $j$ and $k$ that includes node $i$
\\
$C_i^B$			& Betweenness centrality of node $i$
\\
$\lambda$			& Eigenvalue
\\
$x_i$				& Eigenvector
\\
$C_i^E$			& Eigenvector centrality of node $i$
\\
$\alpha_P$			& Attenuation factor
\\
$C_i^P$			& Pagerank centrality of node $i$
\\
\midrule
$\eta$					& Variable neighborhood size
\\
$\mu$				& Genetic algorithm mutation rate
\\
$s$					& Genetic algorithm selection coefficient
\\
$P$					& Population of solutions for the genetic algorithm
\\
$f(i,j)$			& Genetic algorithm fitness function
\\
$\epsilon_B$					& Benefit error from heuristic
\\
$\epsilon_C$					& Cost deviation from heuristic
\\
\midrule
$p$					& Connection probability (Erd\"{o}s-R\'{e}nyi graphs and Klemm and Egu\'{i}lez graphs)
\\
$p_W$				& Rewiring probability (Watts-Strogatz graphs)
\\
$k_L$				& Initial node degree (Watts-Strogatz graphs)
\\
$m_0$				& Initial network size (Barab\'{a}si and Albert graphs and Klemm and Egu\'{i}lez graphs)
\\
$m$					& Degree of new nodes (Barab\'{a}si and Albert graphs)
\\
$p_S$				& node selection probability (Klemm and Egu\'{i}lez graphs)
\\
$p_R$				& Edge removal probability (Delaunay and Voronoi random graphs)
\\
\midrule
$\overline{C_D}$	& Mean degree of a network
\\
$\overline{L}$		& Average path length of a network
\\
$c^w_i$				& Weighted clustering coefficient for node $i$
\\
$w_{ij}$			& Weight of connection between nodes $i$ and $j$
\\
$\overline{C}$		& Weighted clustering coefficient of a network
\\
$\overline{C_r}$	& Weighted clustering coefficient of a completely random network
\\
$\overline{L_r}$	& Average path length of a completely random network
\\
$\gamma$			& Power law exponent
\\
$P(n)$				& Degree distribution
\\
$E$					& Efficiency of a network
\\
$E_r$					& Efficiency of a completely random network
\\
$E_G$					& Global efficiency of a network
\\
$K$					& Number of clusters
\\\bottomrule
\end{tabular}
\caption{List of symbols and their definitions.}
\label{tab:t1}
\end{table}

\section*{Introduction}

Network optimization has generally been focused on solving the following classes of problems (i) finding the shortest path between nodes, (ii) maximizing the flow of information across a network, (iii) minimizing the cost of the flow of information across a network, and (iv) the problems dealing with multiple types of information flows across the network \citep{schrij02,wu04}. One problem, optimizing network connectivity around a specific node with the introduction of new edges, has not been thoroughly explored and yet is important in several domains. Optimizing the network connectivity of additional edges attempts to maximize the number of nodes within a given distance to a focal node to be connected and minimizing the number and length of additional connections is essential in network layout planning for telecommunications and computer systems \citep{resende06,donoso07}, the spread of information or diseases in social networks \citep{eubank04,gavr16}, and the development of neural networks \citep{whitley90}. This network connectivity problem is particularly important with transportation planning in urban environments, where the weights of the network edges can be physical distances or riderships and future street connections or transportation lines can impact flow to established facilities. For example, residential developers could optimize thoroughfare connectivity around existing schools to foster student active commuting and reduce busing costs when planning new developments \citep{line95,auer18,auer20}, and evaluating accessibility and  patient travel time to health care facilities \citep{branas05}.
 
Optimization approaches have been applied to several network problems: the search for new edges that minimize the average shortest path distance in a network \citep{meyerson09}; the minimization of the diameter of the network, i.e. minimizing the the maximal distance between a pair of nodes \citep{demaine10}; and maximization of betweenness centrality \citep{jiang11}. However, the search for new edges, or shortcuts, that maximize connectivity to a focal node and minimize the length of these new edges is less understood and as with the above mentioned graph optimization problems, the search for optimal solutions can become costly when networks are large and complex. This work compares a set of heuristics for this task drawn from a review of combinatorial heuristics \citep{mladenovic07} and from methods used for location models as this problem has many applications where space is an essential component \citep{brimberg11}. In this study, we show that optimization heuristics are preferred for the analysis and practice due to the nonlinearity of the solution space and the optimal solution's dependence on nodal characteristics, such as distance to the focal node.

In connectivity optimization, network nodes are first segmented and assigned to 'close' and 'distant' sets by a specified weighted network distance from the network's focal node. An exhaustive search, where all possible edges from distant to close nodes for a network are evaluated to identify the optimal connections and as a benchmark for the time to find these solutions. This approach ensures that the optimal edges are found. However, as the number of nodes increases and therefore the number of possible connections between close and distant nodes increases, it can become computationally expensive and timely to implement. When the exhaustive search routine was applied to random networks and a real-world street networks, we also discovered that the optimal solution is nonlinearly related to nodal characteristics. To counter this, several heuristics are explored to find the optimal connection utilizing nodal characteristics and possibly in a quicker and less computationally expensive manner: hill climbing with random restart \citep{russell04}; stochastic hill climbing \citep{greiner92}; hill climbing with a variable neighborhood search \citep{mladenovic97}; 
simulated annealing, which has a history of applications in graph problems \citep{kirk83,johnson89,kirkpatrick84}; and genetic algorithms, which has been successfully used for combinatorial optimization \citep{ander94,jara02}. A Tabu heuristic was not employed as it has been observed to not be an effective method for multi-objective optimization problems, whereas simulated annealing and genetic algorithms have shown to be effective \citep{gold86,kim16k}. Among these methods, the genetic algorithm presented here introduces a novel chromosome formulation where the genes are not properties of a specific variable but weights for the probability to move in a given direction in the solution space. This allows the method to dynamically change what solution characteristics to explore while possibly reducing the size of the local neighborhood search.

These optimization heuristics are then applied to randomly generated networks that vary in complexity and size to evauluate their efficacy in finding the optimal connection. Several types of random graph networks were generated to analyze the efficacy of the optimization heuristics for systems with different topologies which are generally representative of naturally occuring and built systems: (1) Erd\"{o}s-R\'{e}nyi networks, (2) Watts-Strogatz networks, (3) Barab\'{a}si and Albert networks, (4) Klemm and Egu\'{i}lez networks, (5) Delaunay triangulation networks, and (6) Voronoi diagrams. Erd\"{o}s-R\'{e}nyi random networks are constructed by randomly creating connection between pairs of nodes with a probability \citep{erdos59}. These networks, even though they have random connections, consistently have short average path lengths and irregular connections, both of which are well found in natural systems. The Watts-Strogatz networks also have random connections but the networks also form clusters, another feature commonly found in real-world networks \citep{watts98}. The Barab\'{a}si-Albert model produces random structures with a small number of highly connected nodes, 'hubs', which are observed in numerous types of networks \citep{bara99,albert02}. Klemm and Egu\'{i}lez networks have random connections, clusters, and hubs \citep{klemm02}.

We also introduce two novel types of random planar network versions of Voronoi diagrams and Delaunay triangulations. The reasons these were considered was that planarity is particularly important in many fields and networks generated from Voronoi diagrams and Delaunay triangles have been used in spatial health epidemiology \citep{john07}, transportation flow problems \citep{steff10,pablo17}, terrain surface modeling \citep{floriani85}, telecommunications \citep{megu01}, computer networks design \citep{liebeherr01}, hazard avoidance systems in autonomous vehicles \citep{anderson12}. Delaunay triangulation maximizes the minimum angles between three nodes, to generate planar graphs with consistent network characteristics \citep{delaun34}. Voronoi diagrams, the dual of a Delaunay triangulation, are composed of points and cells such that each cell is closer to its point than any other point. When edges are randomly removed from the connected Delaunay network or Voronoi network, with weights given by node distance from a focal node, we show that these networks display some of the properties similarly found in the networks mentioned above, such as complexity and randomness, but with the added component of being planar and having edge weights that can be framed as physical distances.

To complement the random network analysis, the network connectivity optimization methods are applied to a study of urban transportation planning. 
We use the network connectivity optimization methods to evaluate the potential costs and benefits of increased thoroughfare connectivity for student active commuting to school. It is assumed that expanding this connectivity around a school would allow for more households, and students, to be included within the walking distance to the school. If more students actively commute to school, this reduces the busing costs for the school system and increases the health and academic achievement of the students \citep{cdc10}. The combinatorial optimization techniques employed here to identify and evaluate new street connections can complement the optimization approaches used for other transportation planning problems, such as greenway planning \citep{line95}, bus stop locations \citep{ibeas10,del12}, and health care accessibility \citep{gu10}.

The following section describes the formulation of the connectivity problem in more detail, the local search methodology, and the optimization heuristics (see Appendix A for the specific pseudocode of the optimization algorithms). The section also details the data used for the study including descriptions of the random networks and the street networks around schools that are used for the transportation study. Results of the heuristics applied to both the random networks and the case study data are also presented in that section. This is followed by a detailed discussion of the heuristic results, the further implications of these techniques for urban transportation planning, and future work for this avenue of research.

\section*{Methods and data}

\subsection*{Formulation of the network connectivity problem}

\begin{figure}[ht]
\centering
\includegraphics[width=\linewidth]{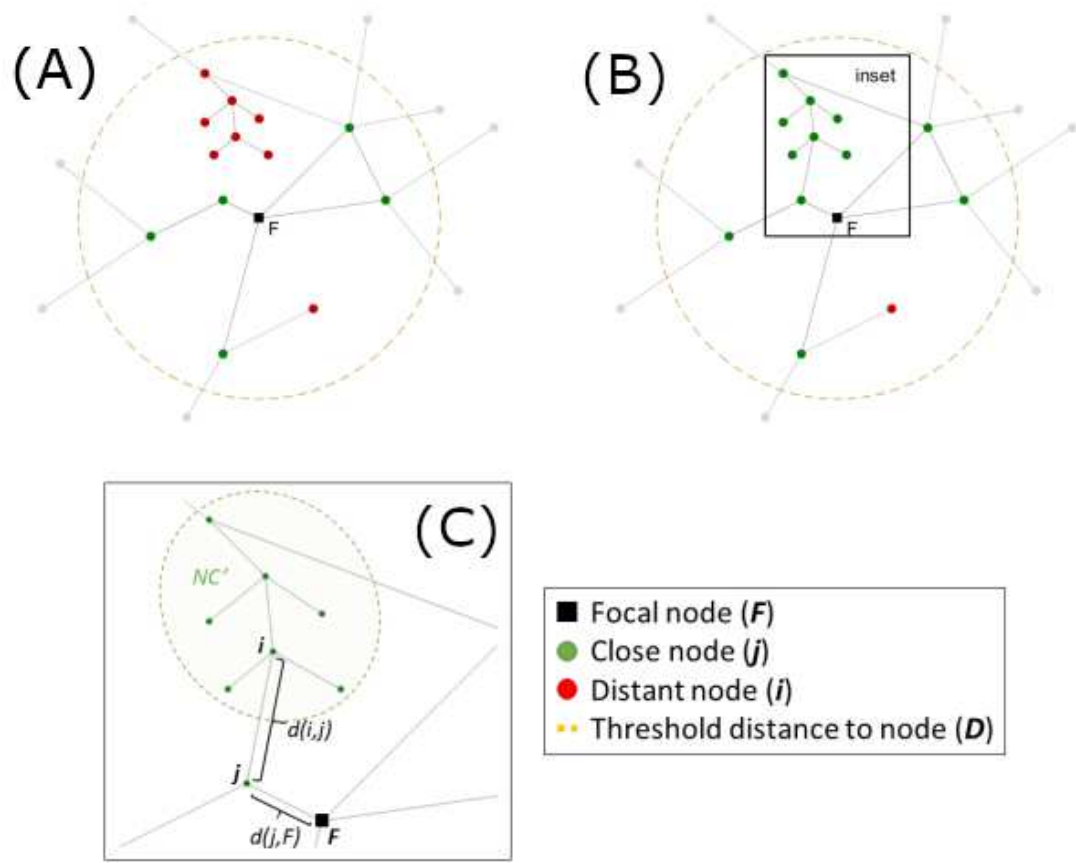}
\caption{Diagram of the network connectivity optimization problem. The close nodes that are within a threshold network distance (orange dashed circle) from the focal node (black square) are colored green, distant nodes that could be within the threshold network distance with additional edges are colored red, and distant nodes that could not be within this distance regardless of additional edges are gray. Figure (A) is an example graph, (B) shows the same graph with the optimal new connection that maximizes the number of additional nodes within the threshold network distance and minimizes the length of the new connection, and the inset (C) highlights this optimal connection, between nodes $i$ and $j$, with the methodological terminology presented in Section 2.}
\label{fig:f1}
\end{figure}

\noindent For the description of the optimization methodology the following nomenclature will be used. The number of nodes is $N$, and nodes are separated into two sets based on their shortest network paths, $d(i,j)$, where $i$ and $j$ are nodes, to the focal node $F$. The nodes that are outside the path distance under consideration, $D$, are assigned to the `distant' set, $N_D \subset N$, i.e. $i \in N_D$ if $d(i,F) > D$.
The nodes that are within this distance are assigned to the `close' set, $N_C \subset N$, i.e. $i \in N_C$ if $d(i,F) \le D$ and $F \in N_C$ (see Figure~\ref{fig:f1} (A)). Node neighborhoods are assigned to the sets $N_i^D$ and $N_j^C$ for the distant and close nodes of $i$ and $j$, respectively by the nodal characteristics described in the following subsection. For a network of size $N$ the number of new connections to evaluate is $\le N^2/4$.

When a new connection is evaluated, any distant nodes that are now within the path distance $D$ to the focal node are assigned to the new set $N_C'$. For example, if a new connection is established between distant node $i$ and close node $j$, then $k\in N_C'$ if $k\in N_D$ and $d(k,i)+d(i,j)+d(j,F) \le D$ (see Figure~\ref{fig:f1} (C)). The number of nodes in $N_C'$ set is considered the benefit of this new connection, $B(i,j)=|N_C'|$. The cost of the new connection is denoted by $C(i,j)$ and for simplicity and this analysis $C(i,j)=d(i,j)$. The optimal solution is the solution with the greatest benefit, or number of new nodes now within the distance to the focal node which can be expressed as the bi-objective function
\begin{equation}
O^* = \max_{(i,j)} (\alpha C(i,j) + \beta B(i,j)),
\end{equation}
where $\alpha$ and $\beta$ are the weights for the costs and benefits, respectively and for this study $\alpha = \infty$ and $\beta = 1$. If $\alpha = \infty$, then the objective is only to minimize costs for the same benefit. Some of the heuristics are also dependent on the number of iterations ($t$) and terminate when the solutions converge, $O^t = O^{t-1}$, or the solution does not improve, $O^{t-1} > O^t$.

\subsection*{Local search methodology}

\noindent The selection of neighboring nodes to improve solutions begins with several evident nodal characteristics (see Supplementary Table S.2 and Figure~\ref{fig:f2}). These nodal characteristics are explored to find the critical network properties for connectivity optimization and their impact on the performance in finding the optimal solution. Nodes are ranked by these characteristics and this creates a multidimensional solution space. A two-level selection process is used, with the following nodal level characteristics: (i) distance to the focal node, (ii) degree centrality, (iii) closeness centrality, (iv) betweenness centrality, (v) eigenvector centrality, (vi) pagerank centrality, (vii) weighted clustering coefficient; and the the following clustering of the characteristics: (i) hierarchical clusters, (ii) network-constrained clusters, and (iii) network modularity. Multiple nodes in a network can have the same degree or assigned to the same cluster, therefore the local searches include a random shuffling routine to evaluate nodes with the same values.

\begin{figure}[ht]
\centering
\includegraphics[width=\linewidth]{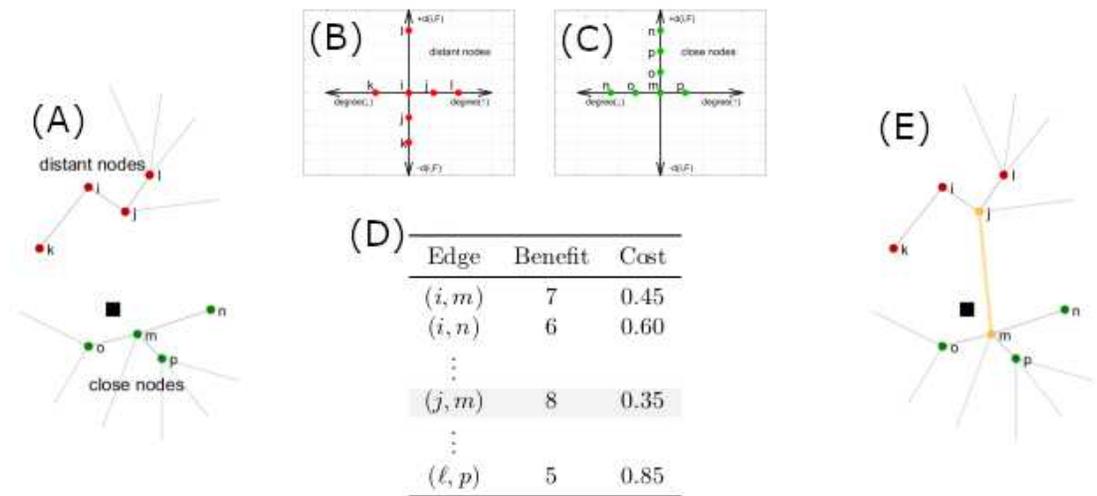}
\caption{Diagram of the local search methodology. Figure (A) shows a generated network with the focal node represented by a black square, the close nodes are colored green and the distant nodes red. Figure (B) shows part of the search space neighborhood for distant node $i$ by increasing and decreasing degree and distance to the focal node, $d(i,F)$ (the other nodal characteristics and clusters are not shown for simplicity). Figure (C) similarly represents the space for the close node $m$. The number of neighbors selected at each iteration of the optimization routine is heuristic dependent. Table (D) gives the costs and benefits for selected connections from the local search and the optimal connection $(j,m)$ for this iteration is shown in (E) an an orange edge.}
\label{fig:f2}
\end{figure}

Nodes are ranked by their distance to the focal node and moving in this solution dimension may result in lower connectivity length costs but may not maximize the number of nodes ultimately connected to the focal node. Ranking and selecting nodes by their centrality, i.e. the importance of the node, could result in maximizing the number of nodes within the specified distance to the focal node but with the possibility of higher connectivity length costs compared to selecting nodes by other characteristics. Several commonly used measures of centrality are explore: degree, the number of edges incident to a node; closeness centrality, the average length of the shortest path between the node and all other nodes in the network \citep{bav50}; betweenness centrality, the frequency of a node included in the shortest paths between all other node pairs \citep{free77}; eigenvector centrality, which is a relative ranking of nodes such that nodes with high values are connected to other nodes with high values \citep{new08}; and pagerank centrality, a variant of eigenvector centrality that ranks nodes based on their probability of being connected to a randomly selected node and which is commonly used in web-page rankings \citep{brin98}.

Nodes are clustered using weighted clustering coefficients, hierarchical clustering, network-constrained clustering, and network modularity. The weighted clustering coefficient of a node is the count of the triplets in the neighborhood of the node and accounts for the weights of the edges times the maximum possible number of triplets that could occur \citep{barrat04}. Nodes are also clustered by their characteristics with hierarchical clustering utilizing Ward's method and the gap criterion. Hierarchical clustering with Ward's method attempts to minimize variance within clusters and maximize variance between clusters \citep{ward63}. The gap criteria is used to identify the optimal number of hierarchical clusters by maximizing the distance between the within-cluster variation and the expected within-cluster variation found from bootstrapping \citep{tib01}. The network-constrained clustering method utilizes the shortest paths between nodes and thereby capturing the network neighbors of each node \citep{yamada06}. Network modularity attempts to cluster nodes by maximizing the number of connections within a cluster and minimizing the number of connections between the clusters. Network modularity accomplishes this by comparing the probability that an edge is in a cluster with the probability a random edge is in the module, i.e. an edge is present in a random graph with the same node degree distribution \citep{newman06}.


\subsection*{Network connectivity optimization heuristics}

\noindent The following techniques were selected for the network connectivity optimization study from their extensive use in optimization (see the Supplementary Material for the algorithms). Parameter selection was simplified for easy comparison of the methods. Random restart, randomly selecting initial nodes to avoid local optima and running the routine until the optimal solution is found, was used for each method to ensure the methods did not converge on suboptimal solutions due to the initial starting values. Six heuristics are employed as below.

\textbf{Exhaustive search (ES).} The exhaustive search optimization routine creates an edge for every combination of distant and close nodes (see Algorithm 1 in the Supplementary Material). Because the results by ES is the optimal, the solution times and the objective values are used to benchmark the solutions by other methods.

\textbf{Hill climbing (HC).} The solution space was observed to be hilly from the exhaustive search results, so several modifications were introduced to the hill climbing technique to address this (Algorithm 2). A stochastic hill climbing {\bf(HCS)}, an advanced search method based on HC, routine is also explored where the selection of nodes for the next iteration is randomly picked with
\begin{equation}
\text{probability}(i,j)=\dfrac{\alpha C(i,j) + \beta B(i,j)}{\sum_{(m,n)}(\alpha C(m,n) + \beta B(m,n))},
\end{equation}
which terminates when a better solution is no longer found (Algorithm 3). A hill climbing algorithm is coupled with a variable neighborhood {\bf(HCVN)} where the size of the neighborhood starts with the nearest neighbors ($\eta=1$) and is updated as follows:
\begin{equation}
\eta =
  \begin{cases}
    1 & \quad \text{if } O^t > O^{t-1}\\
    \eta + 1 & \quad \text{if } O^t \leq O^{t-1}
  \end{cases},
\end{equation}
and the HCVN method terminates after $n_{max}$ is reached (Algorithm 4). 

\textbf{Simulated annealing (SA).} As a meta-heuristic approach, the simulated annealing method randomly selects an initial solution from the solution space to avoid entrapment in a local optima. At each iteration, the heuristic evaluates the neighboring solutions and if it does not find an improved solution, it moves to a new solution with the following probability
\begin{equation}
\text{probability}(i,j) = \exp\left(-\dfrac{O^{t-1}-O(i,j)}{t}\right),
\end{equation}
to obtain an improvement of the solution. The distance of the move decreases with the number of iterations until a better solution is no longer found (Algorithm 6).

\textbf{Genetic algorithm (GA).} The genetic algorithm begins with a population of $P$ randomly selected solutions with a set of chromosomes composed of genes which represent the weights of selecting a neighbor and are all initialized to unity (Algorithm 7). During each iteration of the method, solution scores (fitnesses) are computed by
\begin{equation}
f(i,j)=\dfrac{O(i,j)}{\sum_{(m,n)} O(m,n)},
\end{equation}
and a new generation of solutions are selected based on the following probability condition
\begin{equation}
\text{probability}(i,j) = \dfrac{s*f(i,j) + (1-s)}{\sum_{(m,n)}(s*f(m,n) + (1-s))},
\end{equation}
where $s$ is the selection coefficient. Weak selection, $s \ll 1$, is used to ensure that random mutations impact solution frequency. Crossover is conducted by alternating the weights for the offspring from each parent, also known as cycle crossover \citep{olive87}. Mutations are introduced at a low rate $\mu \ll 1$ for each gene and increase the nodal characteristic or cluster neighbor selection weight by one. The probability that characteristic or cluster $m$ is used to find a neighbor for node $i$ is given by
\begin{equation}
\text{probability(characteristic or cluster)} = \dfrac{\text{gene}(i,m)}{\sum_k \text{gene}(i,k)/K},
\end{equation}
where $K$ is the total number of nodal characteristics and clusters. This formulation ensures that the nodal characteristics or clusters that improve the solution increase in weight, results in a greater probability they will be selected for neighborhood exploration, and reduces the size of the neighborhood search.

\subsection*{Simulated data: Complex random networks}

{\bf Complex random networks.} Several types of random networks were used to evaluate the effectiveness of each heuristic in identifying the optimal new connection. The following types of random undirected networks were generated: (1) Erd\"{o}s-R\'{e}nyi ({\bf ER}) networks, (2) Watts-Strogatz ({\bf WS}) networks, (3) Barab\'{a}si and Albert ({\bf BA}) networks, (4) Klemm and Egu\'{i}lez ({\bf KE}) networks (see Supplementary Figure S.1 (A) -- (D) and for the algorithms used to generate the networks see \cite{prette11}).

{\bf Complex planar networks.} Two novel types of random networks are created here, random Delaunay triangulation  ({\bf DT}) (Supplementary Figure S.1 (E)) and random Voronoi diagrams ({\bf VD}) (Supplementary Figure S.1 (F)). These networks are inherently planar and edges are removed from network nodes randomly based on their distance from the focal node with probability
\begin{equation}
p_R \cdot \dfrac{\max (d(i,F),d(j,F))}{\max_k d(k,F)},
\end{equation}
where $p_R$ is the removal probability and weighted by the normalized edge distance from the focal node.

{\bf Parameter selection.} To compare the efficacy of different optimization methods for different network topologies, identifying the best set of parameters are critical. Parameter values were selected for each type of random network to ensure network complexity (Supplementary Table S.6 summarizes the parameters which were used in the analysis). 
Variation in network size was also explored and the most connected node in each network was selected as the focal node. Uniformly randomly generated edge weights in [0,1] were used for the network distances and the threshold distance was set to ensure that half of the nodes were initially within the distance to the focal node. The costs and benefits were normalized using the ranges from the exhaustive search routine as a benchmark to compare the results from the different optimization methods.

\subsection*{Empirical data: Street networks around schools}

\begin{figure}[ht]
\centering
\includegraphics[width=0.49\linewidth]{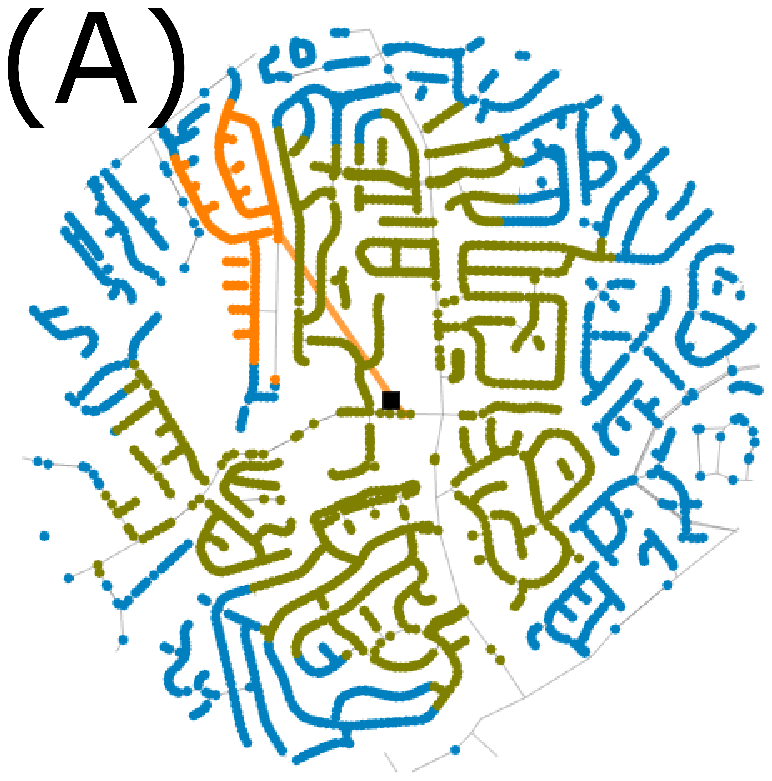}
\includegraphics[width=0.49\linewidth]{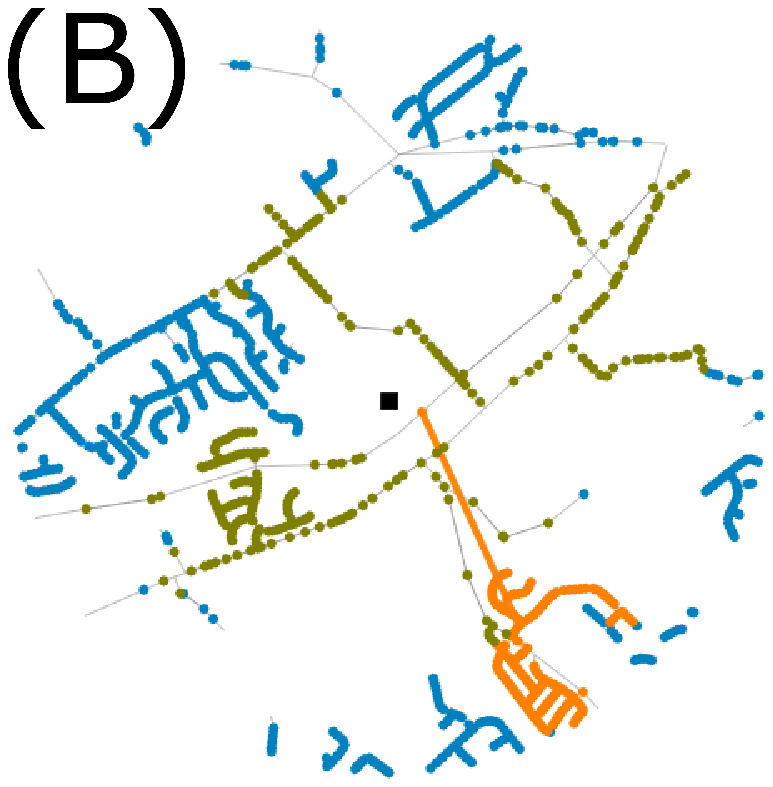}
\caption{Examples of the street networks used for the analysis: (A) a suburban elementary school and (B) a rural elementary school. The blue nodes represent the distant residences, i.e. the residences within the 1-mile Euclidean walking distance but not the 1-mile network walking distance, the green nodes are the close residences within the network walking distance, and the black square represents the school. The orange line denotes the optimal new walking connection that maximizes the number of additional residences (orange nodes) and minimizes the length of the new connection.}
\label{fig:f3}
\end{figure}

\noindent Networks composed of street edges and residence nodes around several schools from a US school system were used for the analysis. Ten suburban and rural schools from Knox County, TN, were selected for the analysis, including seven elementary and three middle, that would benefit the most from increased thoroughfare connectivity, i.e. had the most students within the Euclidean walking distance but not the network distance to the school. Urban schools were not used since the street connectivity around the schools was significantly high and the from additional thoroughfares would be low. Residential nodes were placed on the street networks. The residences within 1 mile and 1.5 miles, for the elementary schools and the middle schools respectively, are considered close nodes while the nodes outside of these distances were classified as distant nodes (see Figure~\ref{fig:f3}). The school networks do not generally display the characteristics of complex network, they had low average degree, large path lengths, and were not efficient, yet have a few intersections (nodes) with a large number of street connections (see Supplementary Table S.4). The networks were evaluated with each optimization method to maximize the number or close residences connected to the school and minimize the distance of the new thoroughfares. The costs and benefits of these street connections were normalized using the ranges from the exhaustive search routine as a benchmark to compare the results from the different optimization methods.

\section*{Results}

\begin{figure}[ht]
\centering
\includegraphics[width=0.49\linewidth]{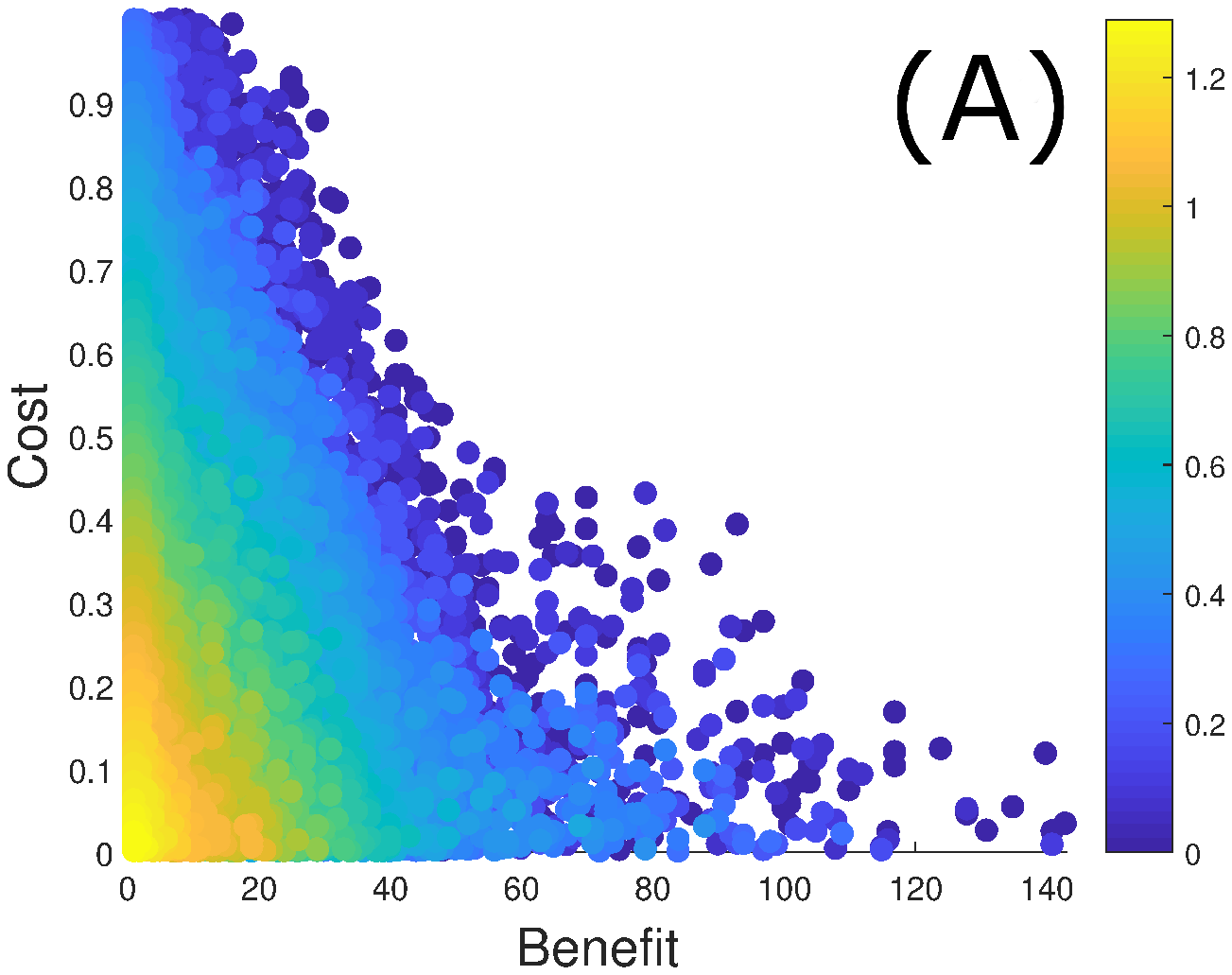}
\includegraphics[width=0.49\linewidth]{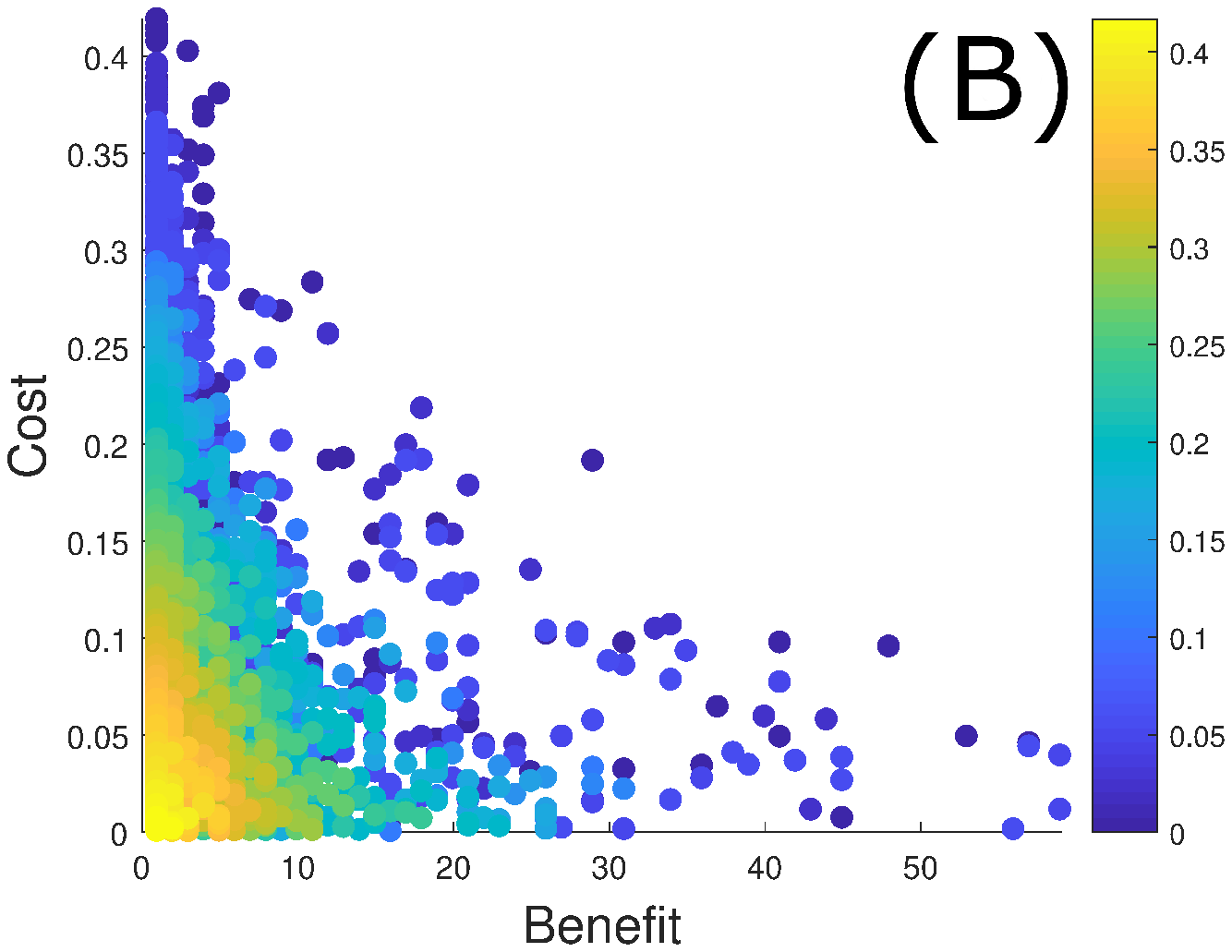}
\includegraphics[width=0.49\linewidth]{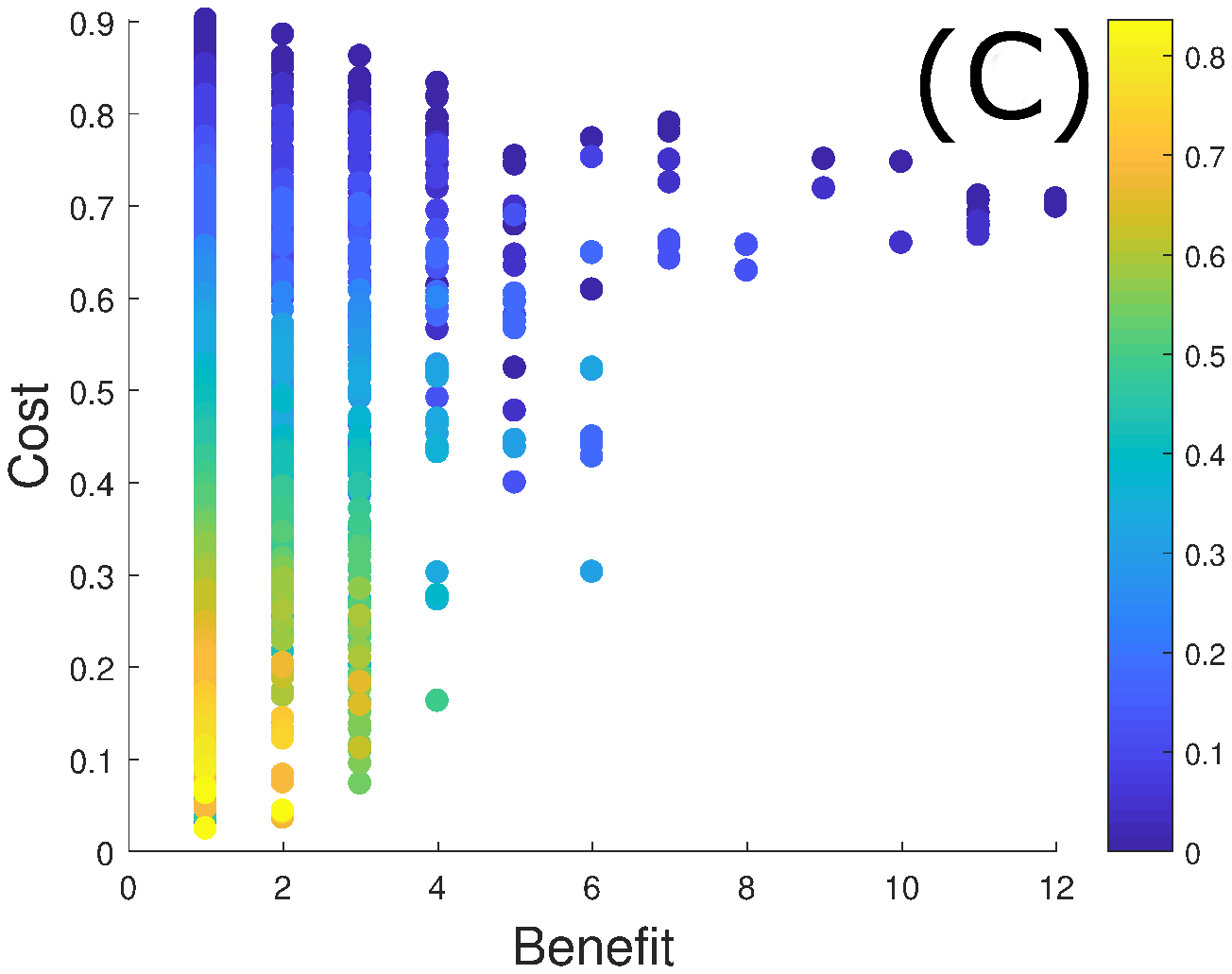}
\includegraphics[width=0.49\linewidth]{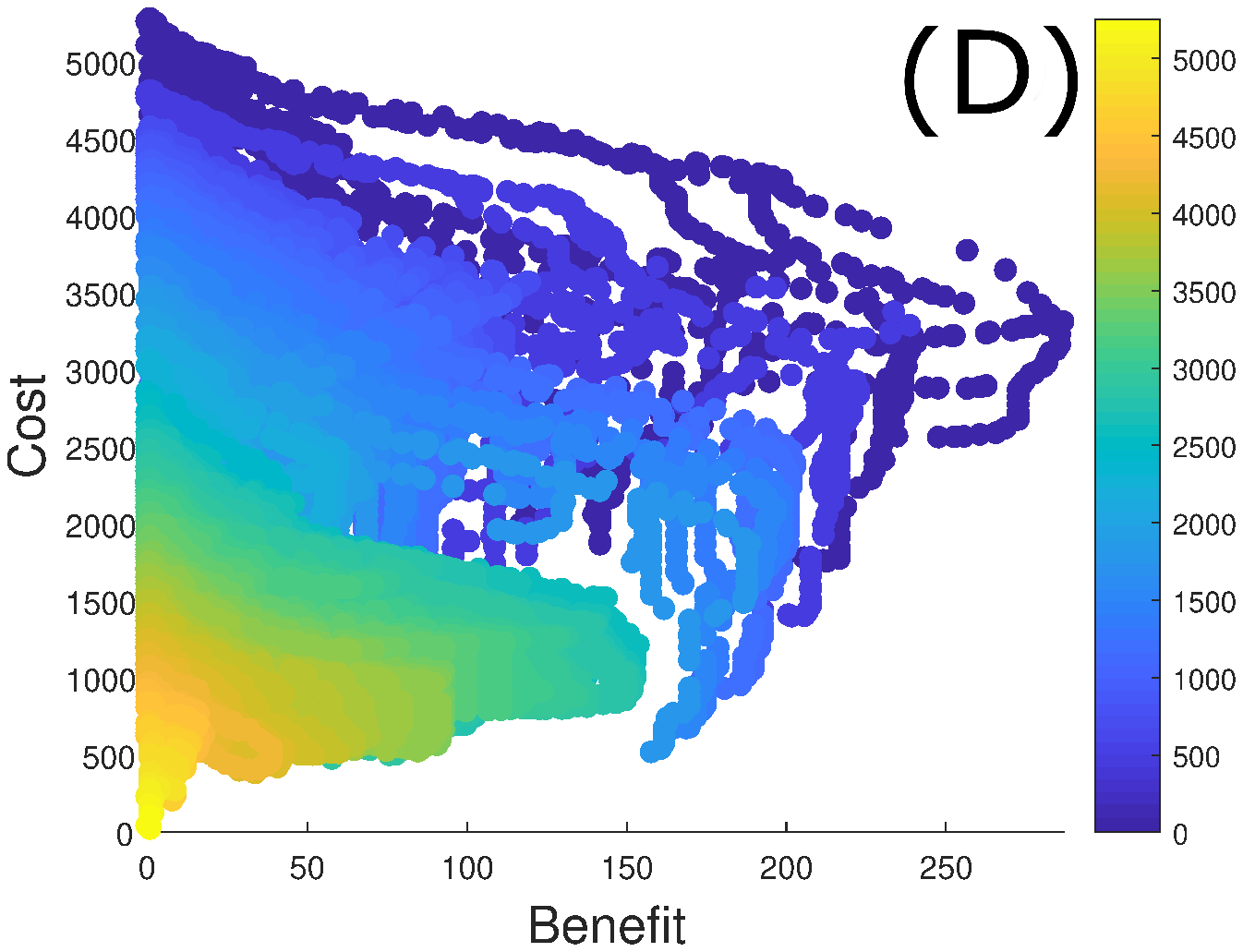}
\caption{The relationship between the distances of the close node and the focal node with the costs and benefits for each solution for different networks. Figure (A) shows the relationship for a Watts-Strogatz network with $N=500$, (B) a Barab\'{a}si and Albert network with $N=500$, (C) a Delaunay network with $N=500$, and (D) a suburban school network ($N\approx4000$). Each point represents a connection between a distant and close node, where the cost is the length of the connection and the benefit is the number of new nodes within the distance to the focal node or school.}
\label{fig:f4}
\end{figure}

Several finding are worthy to note. First, there were consistent nonlinear relationships between the nodal characteristics and the quality of the solutions for each type of random network and the school networks (see Figure~\ref{fig:f4}). There was also significant variation for which nodal characteristics were correlated with the quality of the solution across networks (see Table~\ref{tab:t2}). Among those, the distance between the close node and the focal node and the distance between the distant node and the close node were most often highly correlated with the quality of the solution across networks. The centrality measures were inconsistently related to the solution quality for the random networks. The clustering methods were consistently unrelated to the quality of the solutions for the random networks, while the network modularity for the distant node was correlated with spatial networks and the school networks.

\begin{table}[ht]
\centering
\begin{tabular}{lllrrrrrrr}
\toprule
 &  &          & \multicolumn{7}{c}{Network}                                                                                                                                                       \\
 \cmidrule{4-10}
 &  &          & \multicolumn{1}{c}{ER} & \multicolumn{1}{c}{WS} & \multicolumn{1}{c}{BA} & \multicolumn{1}{c}{KS} & \multicolumn{1}{c}{DT} & \multicolumn{1}{c}{VD} & \multicolumn{1}{c}{Schools} \\
 \midrule
\parbox[t]{2mm}{\multirow{19}{*}{\rotatebox[origin=c]{90}{Nodal characteristics}}} & \parbox[t]{2mm}{\multirow{9}{*}{\rotatebox[origin=c]{90}{close node}}} & $d(j,F)$ & -0.08                  & \textbf{-0.52}                  & \textbf{-0.37}                  & \textbf{-0.48}                  & -0.28                  & \textbf{-0.41}                  & \textbf{-0.30}                       \\
 &  & $C^D_j$  & 0.14                   & 0.08                   & 0.05                   & 0.09                   & -0.01                  & 0.02                   & 0.02                        \\
 &  & $C^C_j$  & 0.10                   & \textbf{0.26}                   & 0.05                   & \textbf{0.15}                   & 0.15                   & 0.15                   & \textbf{0.18}                        \\
 &  & $C^B_j$  & 0.14                   & 0.15                   & 0.05                   & 0.04                   & -0.02                  & 0.03                   & 0.06                        \\
 &  & $C^E_j$  & 0.12                   & 0.11                   & 0.05                   & 0.09                   & 0.01                   & 0.08                   & -0.01                       \\
 &  & $C^P_j$  & 0.14                   & 0.07                   & 0.05                   & \textbf{0.10}                   & -0.03                  & -0.01                  & -0.00                       \\
 &  & $c^w_j$  & 0.00                   & -0.03                  & 0.02                   & 0.00                   & 0.14                   & 0.03                   & *                           \\
 &  & $HC_j$   & 0.01                   & -0.01                  & -0.02                  & -0.01                  & 0.03                   & 0.00                   & -0.00                       \\
 &  & $NM_j$   & -0.01                  & 0.03                   & 0.07                   & 0.06                   & -0.06                  & 0.02                   & \textbf{0.22}                \\
 & & & & & & & & &\\
 &  \parbox[t]{2mm}{\multirow{9}{*}{\rotatebox[origin=c]{90}{distant node}}} & $d(i,F)$ & \textbf{-0.46}                  & \textbf{-0.18}                  & \textbf{-0.12}                  & -0.09                  & -0.07                  & -0.08                  & -0.04                       \\
 &  & $C^D_i$  & 0.17                   & 0.03                   & \textbf{0.08}                   & 0.05                   & 0.29                   & \textbf{0.30}                   & 0.08                        \\
 &  & $C^C_i$  & 0.15                   & -0.01                  & 0.07                   & 0.00                   & \textbf{0.35}                   & 0.21                   & 0.06                        \\
 &  & $C^B_i$  & \textbf{0.19}                   & 0.03                   & 0.07                   & 0.03                   & 0.20                   & 0.20                   & 0.07                        \\
 &  & $C^E_i$  & \textbf{0.21}                   & 0.02                   & 0.07                   & 0.06                   & -0.16                  & -0.01                  & -0.03                       \\
 &  & $C^P_i$  & 0.17                   & 0.01                   & 0.08                   & 0.07                   & \textbf{0.30}                   & \textbf{0.22}                   & 0.01                        \\
 &  & $c^w_i$  & 0.08                   & -0.01                  & 0.03                   & 0.03                   & 0.12                   & 0.12                   & *                           \\
 &  & $HC_i$   & 0.00                   & 0.00                   & 0.02                   & 0.00                   & 0.05                   & -0.03                  & -0.01                       \\
 &  & $NM_i$   & 0.12                   & -0.07                  & 0.02                   & -0.01                  & \textbf{-0.30}                  & -0.19                  & 0.17                        \\ \bottomrule       
\end{tabular}
\caption[Correlation coefficients for the nodal characteristics and the solution benefits for the experimental networks.]{Average correlation coefficients for the random networks (1000 graphs for each network type with $N=1000$) and the ten school networks. The three coefficients with the largest magnitude are highlighted in bold for each network type. (*) There was no variation in clustering coefficients as triplets were not common in the street networks.}
\label{tab:t2}
\end{table}

\begin{figure}[ht]
\centering
\includegraphics[width=0.49\linewidth]{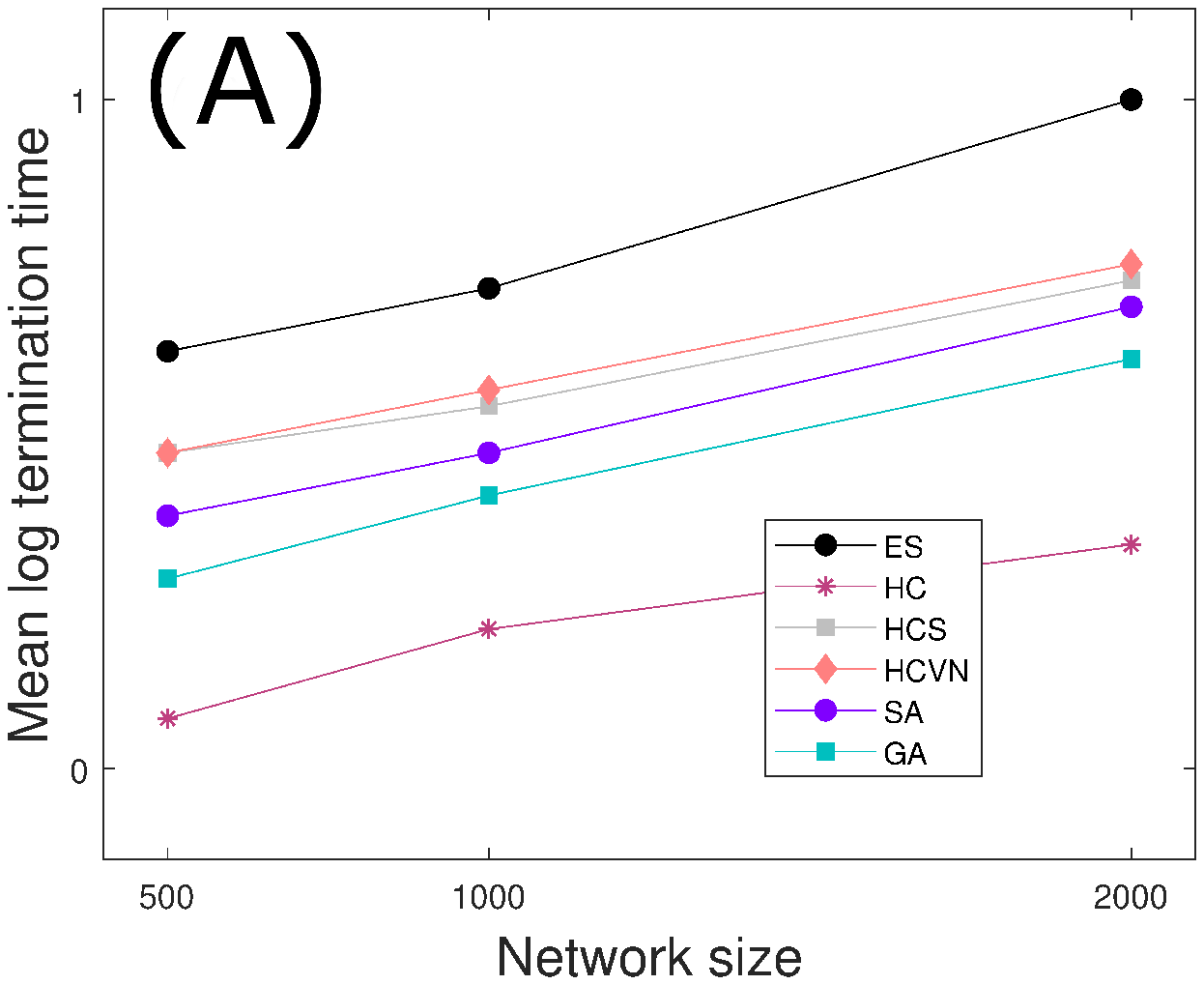}
\includegraphics[width=0.49\linewidth]{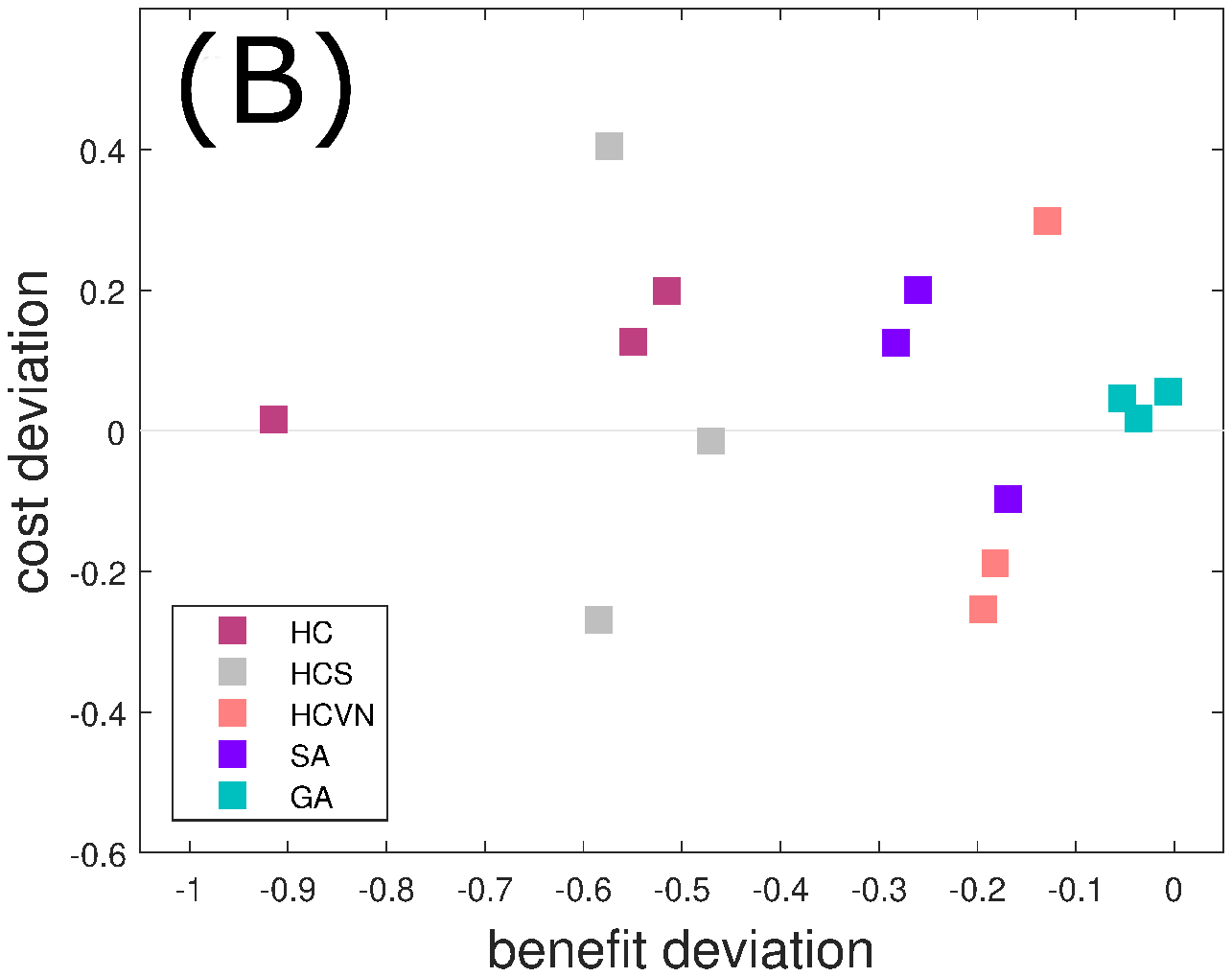}
\includegraphics[width=0.49\linewidth]{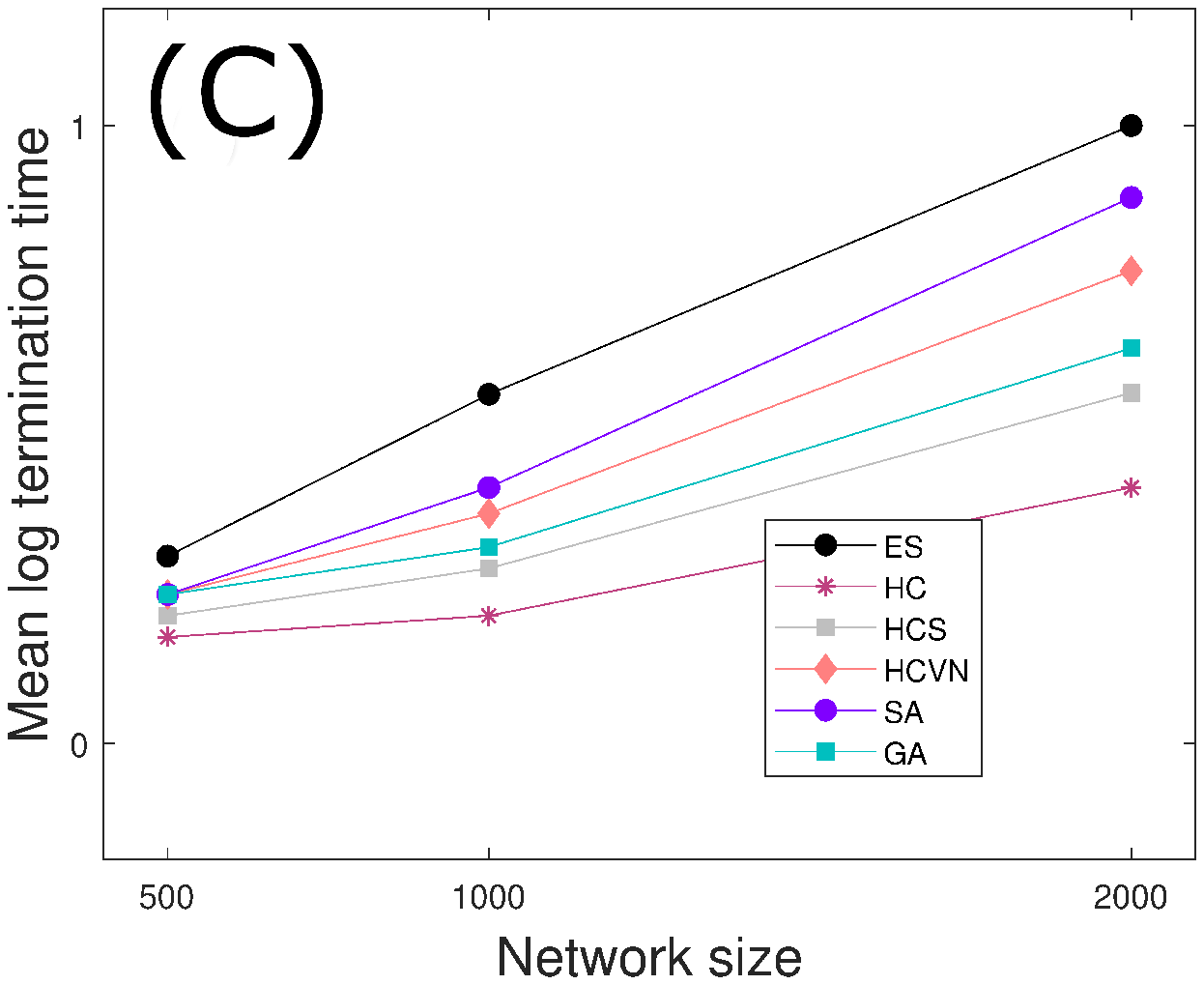}
\includegraphics[width=0.49\linewidth]{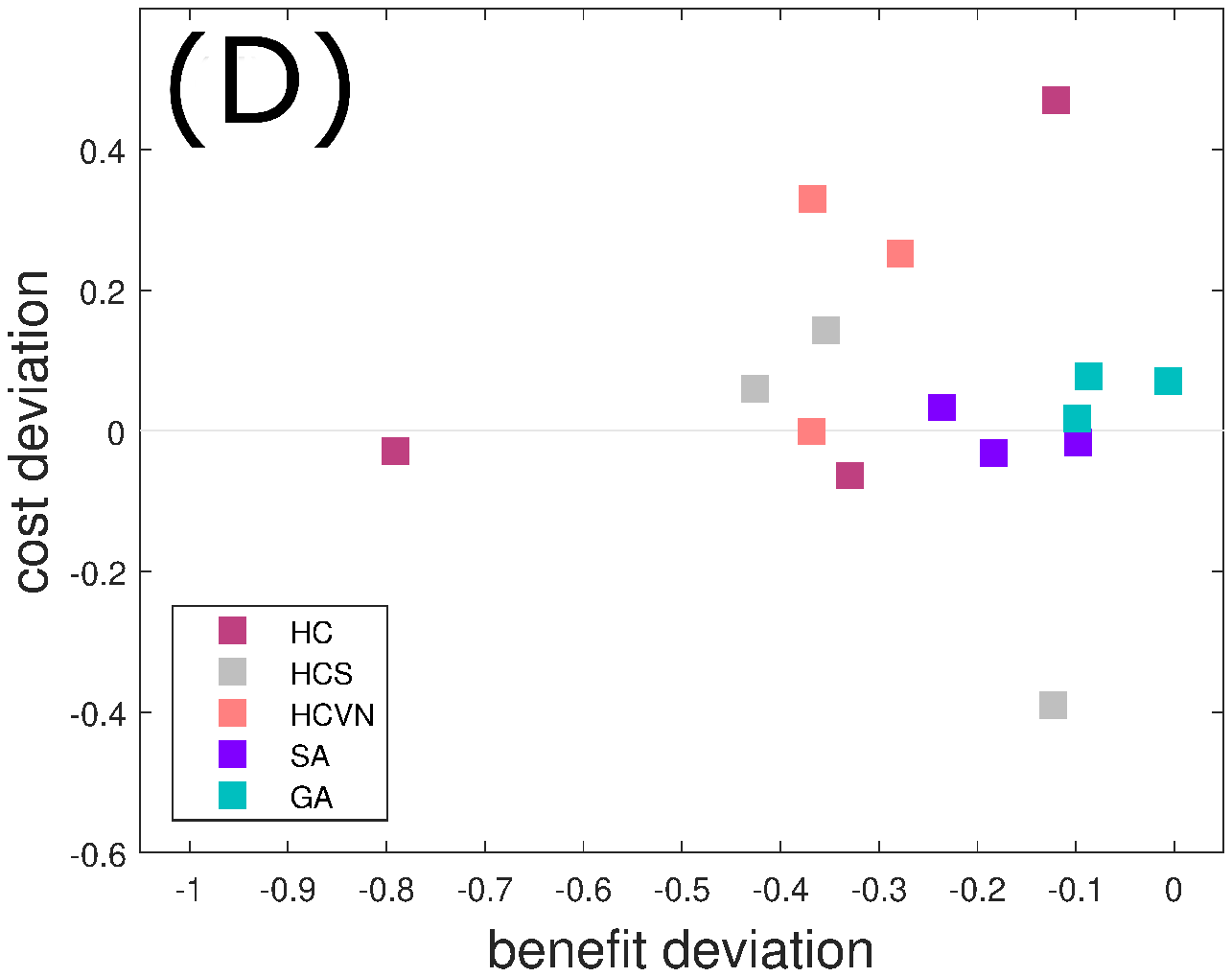}
\includegraphics[width=0.49\linewidth]{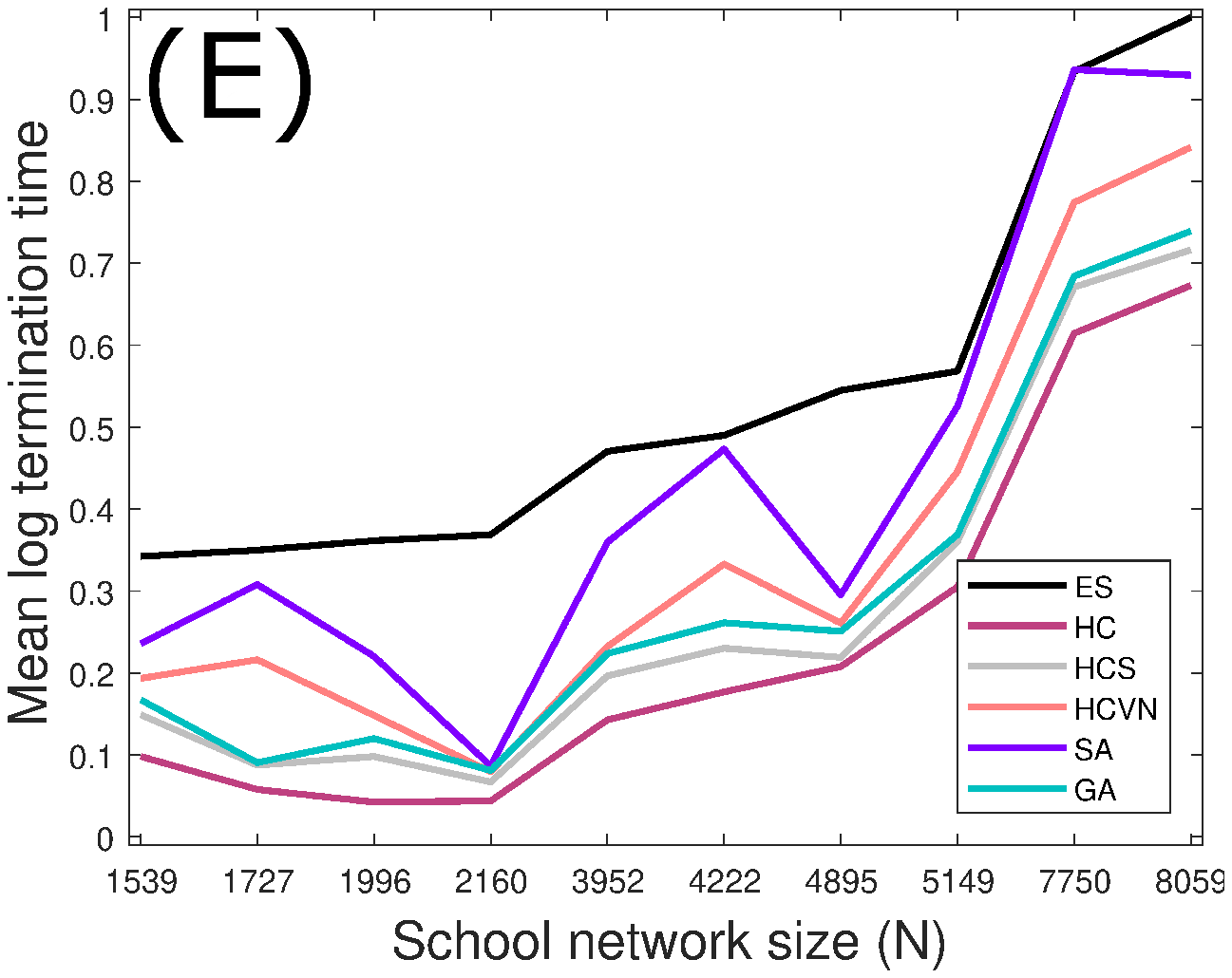}
\includegraphics[width=0.49\linewidth]{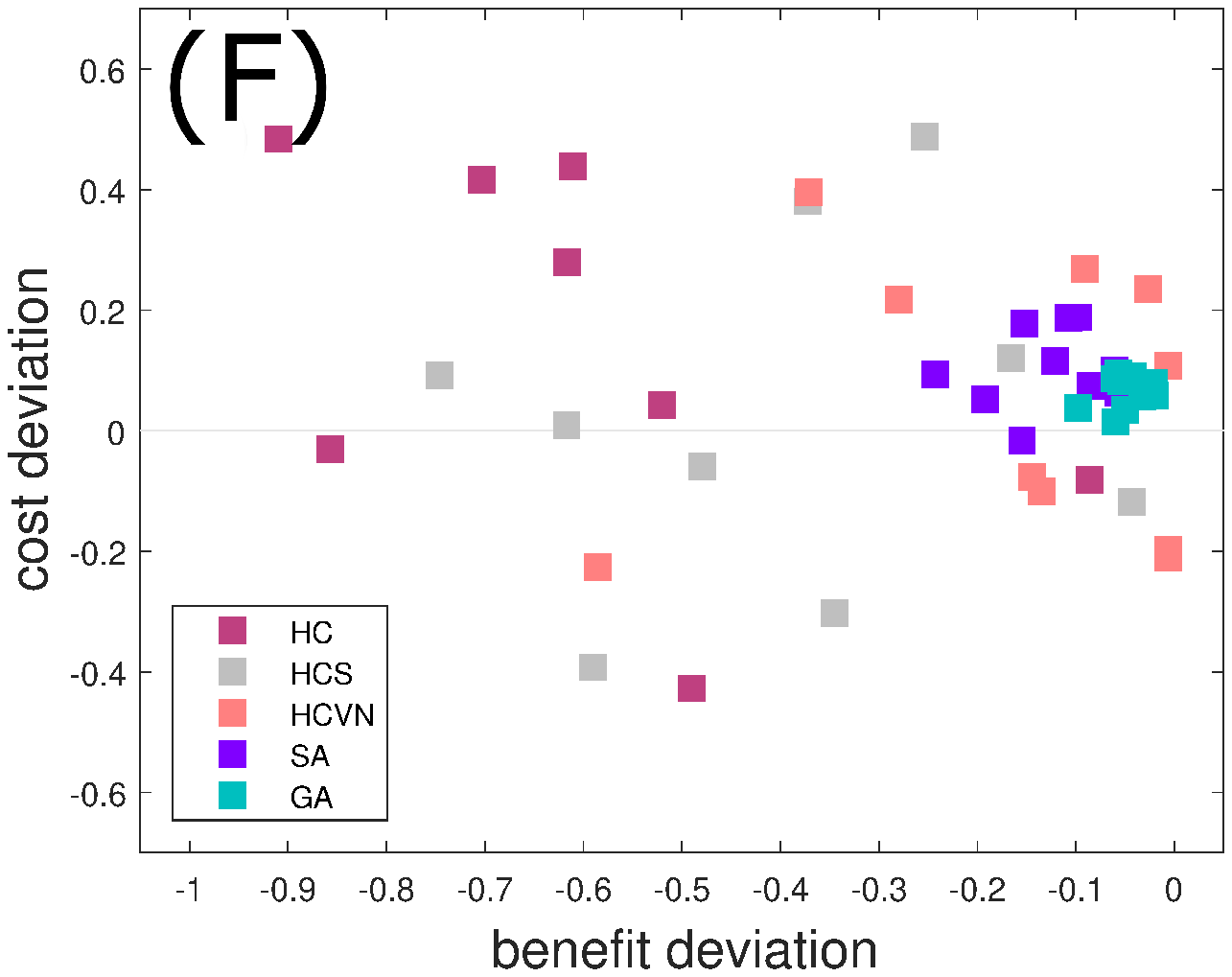}
\caption{The termination times for the heuristics applied to the random networks and school networks: (A) and (B) Erd\"{o}s-R\'{e}nyi networks, (C) and (D) Delaunay networks, and (E) and (F) the ten school networks. These are average times for 1000 random restarts for optimization applied to 1000 random network of each type and size. The times are scaled by the exhaustive search time and log transformed for easier interpretation. The benefits were scaled by the results from the exhaustive search, where a longer connection length is a positive cost deviation and a shorter connection is a negative cost deviation.}
\label{fig:f5}
\end{figure}

Results of the termination times and the optimal solutions deviations from the optimization heuristics applied to the random networks are summarized in Figure~\ref{fig:f5} and Supplementary Figures S.1 and S.2. The hill climbing method was consistently faster for all of the networks, yet had the largest cost and benefit deviations. Simulated annealing and the genetic algorithm had similar termination times, but the genetic algorithm was consistently superior to all of the other methods in approaching the optimal solution. The results from the application of the optimization heuristics applied to the ten school networks are shown in Figure \ref{fig:f5}. The times to termination for each hueristic according to network size consistently followed the following pattern: ES$>$SA$>$HCVN$>$GA$>$HCS$>$HC. The genetic algorithm clearly outperformed the other heuristics, followed by simulated annealing, in terms of cost and benefit deviations (see Figure~\ref{fig:f5} (B), (D), and (F)).

\clearpage

\section*{Conclusion}

\noindent The network connectivity problem introduced in this study is relevant to a wide range of applications and is nontrivial as the number of solutions can become large even for small systems. This type of combinatorial optimization problem highlights the difficulty in determining local search routines a priori. The nodal characteristics were nonlinearly related to the solutions while different characteristics varied in their correlation with solution quality for different networks making it difficult to exclude specific characteristics for network connectivity optimization. Distance to the focal node was consistently related to the quality of the solution as this lowers connectivity length costs, while centrality was intermittently correlated with solution quality it provides greater benefit through more connections. Clustering nodal characteristics did not provide additional useful information from the nodal characteristics for the random networks. This could arise from the following issues: the curse of dimensionality, i.e. large sparse subspaces in the solution space; the nodal characteristics are highly correlated with each other; outliers; finding the appropriate influential nodal characteristics is not possible a priori; and the influence of specific characteristics is dynamic as the heuristics converge. For the school networks: the clustering coefficient was a poor measure due to the lack of triplets in the networks; the network modularity also had poor results possibly due to the measure's inability to account for the spatial component of the nodes; and the network-constrained clusters were also poor in explaining the solutions, due to the complexity of the network topology.

The optimization heuristics save computational time but vary considerably in their ability to find a solution near the optimal. The stochastic hill climbing search was not effective due to the large neighborhood search space explored. In our experiment, the number of solutions checked at each iteration is $>300$ and resulted in a skewed probability distribution of objective values favoring the selection of low values. This degraded the efficiency of the method resulting in the selection of poor solutions. The variable neighborhood search method was similarly not reliable because of the significantly large neighborhood search space (the number of possible solutions explored at a given iteration could be $>5,000$), and had intermediate results with cost and benefit deviations. The simulated annealing heuristic consistently took longer to converge than the other optimization methods from the exploration of suboptimal solutions prior to moving towards better solutions, yet it was able to converge to values close to the optimal solution.

The computational costs and the variance in the importance of nodal characteristics for the random networks and real-world systems highlights the need for a heuristic that is able to quickly and effectively explore the solution space. The genetic algorithm provided in this work offers a solution to this issue and outperformed the other algorithms in terms of the consistently higher solution precision and accuracy. The genetic algorithm is able to dynamically reduce the size of the neighborhood search space and what variables to analyze. This reduction in the local solution search space allows the genetic algorithm presented here to converge on solutions near the optimal in a timely fashion. This shows the power of biologically inspired algorithms to effectively explore multidimensional spaces (commonly found in natural systems) and their potential use in a wide variety of disciplines, including specific applications for planning and health care.



Application of these methods and heuristics to multi-level networks, such as telecommunication systems, higher dimensional real-world networks (transportation networks with elevation), directed networks, and additional planar random networks (e.g. Gabriel graphs) should be conducted. Different distance measures to the focal node, such as the Hamming distance, could also be evaluated for different applications, and other real world examples should be used for analysis. The methods presented here do not evaluate whether the new connections intersect existing edges and attempts to incorporate such a feature resulted in unrealistic computational times. Optimizing this feature is currently being developed as is a tool for ArcGIS and Python for planners and researchers to utilize.

\section*{Acknowledgments.} We would like to thank Alex Zendel (GIS Analyst at the Knoxville-Knox County Metropolitan Planning Commission) for providing the street networks and residential data around the schools.

\end{document}